\documentclass[a4paper,12pt]{article}

\usepackage{amssymb}
\usepackage{dsfont}
\usepackage{color}
\usepackage[utf8]{inputenc}
\usepackage{graphicx}
\usepackage{epstopdf}
\usepackage{hyphenat}
\usepackage{amsmath}
\usepackage{wrapfig}
\usepackage{empheq}
\usepackage{textcomp}
\usepackage{subfigure}
\usepackage{rotating}
\usepackage{wrapfig}
\usepackage{pdfpages}
\usepackage{color}
\usepackage{amsmath}
\usepackage{amsfonts}
\usepackage{verbatim}
\usepackage{amssymb}
\usepackage{graphicx}
\usepackage{amsmath,amssymb,calc}
\usepackage{setspace}
\usepackage{color}
\usepackage{amsmath}
\usepackage{amsfonts}
\usepackage{verbatim}
\usepackage{amssymb}
\usepackage{graphicx}
\usepackage{array}
\usepackage{caption}
\usepackage{epstopdf}
\usepackage[pdf]{pstricks}
\usepackage{upgreek}
\usepackage{cmll}
\usepackage{latexsym}
\usepackage{braket}
\usepackage{cite}
\usepackage{amsmath,amssymb,amsfonts,graphicx}
\usepackage{epsfig}
\usepackage[left=2.4cm,top=3.3cm,right=2.4cm,bottom=3.3cm,bindingoffset=0cm]{geometry}

\newcommand{\beq}{\begin{eqnarray}}
\newcommand{\eeq}{\end{eqnarray}}
\newcommand{\bea}{\begin{eqnarray}}
\newcommand{\eea}{\end{eqnarray}}
\newcommand{\be}{\begin{equation}}
\newcommand{\ee}{\end{equation}}

\def\1{\mathbbm{1}}

\def\tr{\mathrm{tr}}

\def\nn{\nonumber}

\numberwithin{equation}{section}

\newcommand{\hA}[0]{\widehat{A}}
\newcommand{\hF}[0]{\widehat{F}}
\newcommand{\mF}[0]{\mathcal{F}}
\newcommand{\mA}[0]{\mathcal{A}}
\newcommand{\mU}[0]{\mathcal{U}}

\begin{document}

\title{
\begin{flushright}\ \vskip -1.5cm {\small {IFUP-TH-2017}}\end{flushright}
\vskip 40pt
\bf{ \Large  From the Sakai-Sugimoto Model to the \\ Generalized Skyrme Model}
\vskip 18pt}
\author{{\large Lorenzo Bartolini, Stefano Bolognesi and Andrea Proto} \\[20pt]
{\em \normalsize
Department of Physics ``E. Fermi'', University of Pisa
  and INFN Sezione di Pisa}\\[0pt]
{\em \normalsize
  Largo Pontecorvo, 3, Ed. C, 56127 Pisa, Italy \footnote{emails:  lorenzobartolini89@gmail.com, stefano.bolognesi@unipi.it, proto.andrea@yahoo.com}}\\[3pt]
\\[10pt]}
\vskip 10pt
\date{November 2017}
\maketitle
\vskip 0pt

\begin{abstract}

We derive the generalized Skyrme model as a low-energy effective model of the Sakai-Sugimoto model. The novelty with the past is the presence of the sextic term  equal to the topological charge squared.
This term  appears when the $\omega$ meson, and the tower of states on top of it, are integrated out.
We claim that, in the small 't Hooft coupling limit, the instanton is well described by a Skyrmion arising from the  low energy effective Lagrangian of the Sakai-Sugimoto model.
The sextic term plays a dominant role in this limit.  Moreover, when a pion mass term is added we recover the BPS Skyrme model in the small 't Hooft coupling limit.

\end{abstract}

\newpage


\section{Introduction}

The Sakai-Sugimoto (SS) model is a top-down holographic model of QCD  \cite{Sakai:2004cn,Sakai:2005yt}.
The model, among all the holographic attempts, is one of the closest example to QCD  and being a top-down model it has very few parameters to adjust.
The SS model has been successfully  applied to reproduce qualitative or semi-quantitative properties of QCD both in the mesonic and in the baryonic sector.
The mesonic sector contains the pions, which are the Goldstone bosons of the chiral symmetry breaking, plus a tower of massive vector bosons among which the $\omega$-meson and the $\rho$-mesons.
The  Lagrangian, restricted to the pions fields, reproduces exactly the Skyrme model.
The model incorporates all the features of the large-N expansion and in particular the identification between baryons and Skyrmions-Instantons, which are solitonic objects made out of the mesonic fields  \cite{Hong:2007kx,Hata:2007mb,Hashimoto:2008zw}.

In this paper we want to explore more in detail the relation between the Sakai-Sugimoto model and the effective Skyrme model.
Our first task is to find a missing important piece in the  pions effective Lagrangian: the sextic term.
We just said that, restricting to the pions fields, just by setting the vector mesons to zero, we obtain the familiar Skyrme model with the two terms which are most commonly considered, the quadratic and the quartic one.
However the correct way to obtain a low-energy effective action is not to set to zero the massive fields by brute force. We should  instead integrate them  out in order to find any residual interactions to be included in the derivative expansion of the low-energy effective action.
We find in fact that a new term is generated by this procedure: a sextic term in number of derivatives which correspond the the topological charge squared.
This is analogue to what happens in the Skyrme model coupled to the $\omega$ vector meson \cite{Jackson:1985yz} where integrating out the massive meson generates a term proportional to  the topological charge squared. Our mechanism  is a generalization of the one just cited adapted to the Sakai-Sugimoto model where the whole tower of vector mesons enters into play. The sextic term has also been previously discussed in  \cite{Colangelo:2012ipa}.

We thus obtain the so called ``generalized Skyrme model" which is the model that contains four terms in the Lagrangian ${\cal L} = {\cal L}_0 +  {\cal L}_2+ {\cal L}_4 +  {\cal L}_6$, where the subscript corresponds to the number of derivatives of the pion field.  ${\cal L}_0$ is the mass term, or a generic potential,  which is generated by an explicit breaking of the chiral symmetry,  ${\cal L}_2$ is the quadratic Dirichlet term,  ${\cal L}_4$ is the quartic Skyrme term and ${\cal L}_6$ is the topological charge squared term.
The generalized Skyrme model  is the most generic effective model with the requirement of having at most two time derivatives.
By setting ${\cal L}_6$ to zero we re-obtain the ordinary massive Skyrme model which has been studied in depth in the past and has the  drawback of predicting too large classical binding energies for nuclei. By setting ${\cal L}_2$ and ${\cal L}_4$ to zero we obtain the so called BPS Skyrme model \cite{Adam:2010ds,Adam:2013wya}. It is a model in which Skyrmions, in any topological sector, saturate a Bogomolny bound and have a infinite dimensional moduli space corresponding to the volume preserving diffeomorphisms. Starting from the BPS Skyrme model and treating ${\cal L}_2$ and ${\cal L}_4$ as perturbations, it  is conjectured to have small nuclear binding energy and to be very close to the phenomenology of liquid-drop model \cite{Adam:2013wya,Gillard:2015eia}.
The interesting aspect of the low-energy effective action of the SS model, for small 't Hooft coupling,  is that it falls exactly into this class with the coefficient of all the terms precisely fixed by the UV theory.

In the second part of this paper we consider the baryonic sector of the SS model and how the generalized Skyrme model may help to understand it.
The picture at large 't Hoof coupling $\lambda$ is the one that received most of the attention in the past.
The whole string theory set-up is in fact well approximated by semi-classical computation only in this limit, together with the limit of  large $N_c$.
When the 't Hooft coupling is large the SS model has instantons solutions, very well approximated by small BPS instantons whose size scales as ${\cal O}(1/\sqrt{\lambda})$, which is much smaller than the bulk radius of curvature.
These instantons correspond to the baryons of the dual QCD.
They carry the same quantum number of the Skyrmions but they are quite different from the Skyrmions that would be obtained by solving the Skyrme effective Lagrangian in isolation. When instantons are very small they probe deep into the fifth holographic dimension and all the tower massive vector mesons enter in their structure.

We here focus our attention the another, much less studied, limit of {\it small} 't Hooft coupling. This limit lies outside the applicability of the top-down string holographic model, but it does make sense if we consider our model as a bottom-up phenomenological model.
Moreover, when we calibrate the SS model to the real QCD we have to choose a particular 't Hooft coupling which in general is never too large or to small, it falls in the middle between  the two limits.

We claim that in the  small 't Hooft coupling limit the instanton becomes very large and eventually can be studied by considering only the generalized Skyrme effective obtained by integrating out all the massive vector meson. So in this limit the instanton really turns out to be a Skyrmion. In particular we find that at very small 't Hooft coupling only few terms in the effective Skyrme Lagrangian are important. If pions are massless they are $ {\cal L}_2 +  {\cal L}_6$. If pions are massive they are instead $ {\cal L}_0 +    {\cal L}_6$ thus reproducing the result of the BPS Skyrme model. We thus find the 't Hooft coupling interpolates between to distinct BPS models for baryons: small self-dual instantons for large $\lambda$ and large BPS Skyrmions for small $\lambda$. This is a generalization of what happens in lower dimensions with the baby-Skyrme model \cite{Bolognesi:2014ova}.

The paper is organized as follows. In Section \ref{due} we discuss the pion effective Lagrangian obtained by integrating out the massive vector meson. In Section \ref{tre} we consider the baryon at small 't Hooft coupling which become the Skyrmions of the generalized Skyrme model. We conclude in Section \ref{quattro}.

\section{The sextic term derivation from the SS model}
\label{due}

The model, after dimensional reduction to an effective $5D$ theory, is described by a Yang-Mills/Chern-Simons action \cite{Sakai:2005yt}:
\bea\label{SYMCS}
 && S= S_{YM} + S_{CS}   \phantom{ \frac{1}{2}}\nonumber \\
&&   S_{YM} =  -\kappa \text{tr}\int d^4 xdz \left[ \frac{1}{2}h(z)\mathcal{F}_{\mu \nu}^2 + k(z) \mathcal{F}_{\mu z}^2 \right]  \nonumber \\
&& \  S_{CS} =  \frac{N_c}{384 \pi^2} \epsilon_{\alpha_1 \alpha_2 \alpha_3 \alpha_4 \alpha_5} \int d^4 xdz \hA_{\alpha_1} \left[ 6\tr\left(F_{\alpha_2 \alpha_3}^a F_{\alpha_4 \alpha_5}^a \right)+2\tr\left(\hF_{\alpha_2 \alpha_3}\hF_{\alpha_4 \alpha_5}\right)\right]
\eea
where $\kappa \equiv a N_c \lambda$ with $a\equiv (216\pi^3)^{-1}$, and $k(z)= (1+z^2)$, $h(z)=k(z)^{-1/3}$.
The field content is given by a $U(N_f)$ connection $\mathcal{A}$:
\begin{equation}\label{fielddef}
  \mathcal{A}= \hA \frac{\mathds{1}}{N_f} + A^a T^a
\end{equation}
where $T^a$ are the generators of $SU(N_f)$ normalized to obey $\tr(T^a T^b)=\frac{1}{2}\delta^{ab}$.
We will work in the $N_f=2$ case, thus accounting for the up and down quarks: in this case $T^a = \frac{\tau^a}{2}$.

We can go to the $\mA_z=0$ gauge with the transformation
\begin{equation}\label{gaugeA}
\begin{split}
  \mA& \rightarrow \mA^g = g^{-1}\mA g + ig^{-1}dg\\
  g&\equiv \exp\left(i\int_{0}^{z}dz^{\prime}\mA_z(x,z^{\prime})\right) \ .
  \end{split}
\end{equation}
In this gauge the Chern-Simons Action assumes the form:
\bea\label{Scsgauge}
  S_{CS}&=&\frac{N_c}{384\pi^2} \epsilon^{\mu_1 \alpha_2 \cdots \alpha_5}\int d^4 xdz\hA_{\mu_1}\left[6\tr\left(F_{\alpha_2\alpha_3}F_{\alpha_4\alpha_5}\right)+2\tr\left(\hF_{\alpha_2\alpha_3}\hF_{\alpha_4\alpha_5}\right)\right] \nn \\
  &=& \frac{N_c}{96\pi^2} \epsilon^{\mu_1 z \mu_3 \mu_4 \mu_5}\int d^4 xdz\hA_{\mu_1}\left[6\tr\left(F_{z\mu_3}F_{\mu_4\mu_5}\right)+\hF_{z\mu_3}\hF_{\mu_4\mu_5}\right] \ .
\eea
In particular we shall need the following equation of motion for the $U(1)$ $\hA_{\mu}$ field:
\begin{equation}\label{eomAmu}
  -\kappa\left( h(z)\partial_{\nu}\hF^{\mu\nu}+\partial_z \left(k(z)\hF^{\mu z}\right)\right) + \frac{N_c}{64\pi^2}\epsilon^{\mu \alpha_1 \cdots \alpha_4} \tr\left(\mF_{\alpha_1 \alpha_2}\mF_{\alpha_3\alpha_4}\right)=0
\end{equation}


In the gauge adopted, we know that the following fields expansion holds \cite{Sakai:2004cn}:
\begin{equation}\label{modesakai}
  \mathcal{A}_{\mu}= \mU^{-1}\partial_{\mu}\mU\psi_+ + \sum_{n=1}^{\infty}B_{\mu}^{(n)}(x)\psi_{n}(z)
\end{equation}
where the first term accounts for the pion component $\psi_+ \equiv -\frac{i}{2}\left(1+\frac{2}{\pi}\arctan z\right)$,
while the second sum includes all vector and axial-vector mesons.
We employ an Abelian ansatz for the vector meson part and more over we assume that it factorizes as follows:
\begin{equation}\label{decomposition}
  \mA_{\mu} =
\left\{
  \begin{aligned}
 \hA_{\mu} & = B_{\mu}(x) \chi (z)\\
 A_{\mu} & = \mU^{-1}\partial_{\mu} \mU \psi_+ (z) \\
 \end{aligned}
\right.
\end{equation}
With this ansatz  the field strength becomes
\begin{equation}\label{expansion}
\begin{split}
  F_{\mu \nu}&=  \left[R_{\mu} , R_{\nu} \right] \psi_+ \left(i\psi_+ -1\right) \\
  F_{z\mu}&=R_{\mu}\psi_+ ' \\
  \hF_{\mu \nu}&= f_{\mu\nu}\chi 
\\
  \hF_{ z\mu}&= B_{\mu}\chi'
  \end{split}
\end{equation}
where we have defined $R_{\mu} = \mU^{-1}\partial_{\mu}\mU$ and $f_{\mu \nu } \equiv \partial_{\mu}B_{\nu}-\partial_{\nu}B_{\mu}$.\

Plugging (\ref{expansion}) and (\ref{decomposition}) into (\ref{eomAmu}) we obtain the following differential equation:
\bea
\label{equ}
&&  2\kappa zB_{\mu}\chi' + \kappa(1+z^2)B_{\mu} \chi''
+\kappa h(z)\partial_{\nu}f^{\nu \mu}\chi + \nonumber
 \\
&&   +\frac{N_c}{16\pi^2}\epsilon^{\mu z \mu_1 \mu_2 \mu_3}\left\{\tr\left(R_{\mu_1}\left[R_{\mu_2},R_{\mu_3}\right]\psi_+\psi_+ ' (i\psi_+ -1)\right) + \frac{1}{2} \left(B_{\mu_1} f_{\mu_2\mu_3}\chi'\chi\right)\right\}=0 \ .  
\eea
Solving this we get $B_{\mu}$ as a function of $R_{\mu}$.
We want to obtain a low energy $4D$ effective action so we want to keep the term with lowest number of four-dimensional derivatives.
For this reason we omit both  terms in (\ref{equ}) that depend on $f_{\mu \nu}$.
Keeping the remaining terms, we obtain two equations, one for $B_{\mu}$ and the other for the profile function $\chi(z)$ in the holographic direction:
\bea
  & & 2z\chi' + k(z) \chi''=\frac{N_c}{16\kappa \pi^2}\psi_+\psi_+ ' (i\psi_+ -1)
\label{eqfchi} \\
   & & B_{\mu}(x) =- \epsilon_\mu^{\quad z \nu_1 \nu_2 \nu_3}\tr\left(R_{\nu_1}\left[R_{ \nu_2},R_{ \nu_3}\right]\right) \label{eqB} \ .
  \eea
This justifies the factorization of the ansatz  (\ref{decomposition}) for the vector meson part.
The function $\chi(z)$ can be obtained in an analytical form by solving its equation of motion with boundary conditions $\chi(\pm \infty)=0$:
\begin{equation}\label{solutionchi}
  \chi = -\frac{N_c}{64\pi^3 \kappa}\left(\frac{5\pi^2}{48}-\frac{1}{2}\arctan^2(z)+\frac{1}{3\pi^2}\arctan^4(z)\right) \ .
\end{equation}
It is trivial to check that with this solution for $B_{\mu}$, the neglected terms in (\ref{equ}) are all of a higher $L^{-1}$ order than the ones we kept.

We now plug the ansatz (\ref{decomposition}) into the  action and use equations (\ref{eqfchi}), (\ref{eqB}) to express everything in terms of $R_{\mu}$.
For the Chern-Simons term we obtain
\begin{equation}\label{actionansatz}
S_{CS}=\frac{N_c}{96 \pi^2} \epsilon^{\mu_1 z \mu_3 \mu_4 \mu_5} \int d^4 xdz B_{\mu_1}\chi \left[6\tr\left(R_{\mu_3}\left[R_{\mu_4},R_{\mu_5}\right]\right)\psi_+\psi_+'(i\psi_+-1) + B_{\mu_3}f_{\mu_4\mu_5}\chi'\chi\right] \ .
 \end{equation}
Let us first concentrate on the first term. If we make use of the equation for $B_{\mu}(x)$ and use the fact that a commutator antisymmetrized via the totally antisymmetric tensor amounts to two times the product, we find
\bea\label{actionsextic}
  S_{CS}^{(1)} &=& -\frac{N_c}{4 \pi^2} \epsilon^{ z \mu_1 \mu_3 \mu_4 \mu_5}\epsilon^{z \mu_1 \nu_3 \nu_4 \nu_5} \int d^4 x \tr\left(R_{\mu_3}R_{\mu_4}R_{\mu_5}\right) \tr \left(R_{\nu_3}R_{\nu_4}R_{\nu_5}\right)\int dz\chi \psi_+\psi_+'(i\psi_+-1)  \nonumber \\
  & =& -\frac{N_c}{4 \pi^2} \int d^4 x\left(\epsilon^{z \mu_1 \nu_3 \nu_4 \nu_5} \tr \left(R_{\nu_3}R_{\nu_4}R_{\nu_5}\right)\right)^2\int dz\chi \psi_+\psi_+'(i\psi_+-1) \ .
  \eea
Performing the integral and using the solution for $\chi(z)$  (\ref{solutionchi}), we obtain the following sextic term:
\begin{equation}\label{sextic}
  S^{CS}_{6}= \frac{51 N_c}{4480 \lambda} \int d^4 x\left(\epsilon^{z \mu_1 \nu_3 \nu_4 \nu_5}\tr \left(R_{\nu_3}R_{\nu_4}R_{\nu_5}\right)\right)^2 \ .
\end{equation}
The second term of (\ref{actionansatz}) vanishes because it can be reduced to a boundary term and $\chi(z)$ vanishes for $z\rightarrow\pm \infty$:
\begin{equation}
\begin{split}
S_{CS}^{(2)}\propto &\int_{-\infty}^{+\infty} dz \chi ' \chi^2 = \frac{1}{3} \int_{-\infty}^{+\infty} dz \partial_z\left(\chi^3\right) = \frac{1}{3}\left[\chi^3\right]^{+\infty}_{-\infty} = 0 \ .
\end{split}
\end{equation}

Another contribution to the sextic term arises from the Abelian part of the Yang-Mills action.
Of the two terms, the one without derivatives in $z$ turns out to be of order $L^{-8}$, hence we neglect it, as we have done with its contribution to the equation of motion. Then, the only contribution to a sextic term comes from the action:
\begin{equation}\label{sexticym}
\begin{split}
  S^{YM}_{6}= &-\frac{\kappa}{2} \int dzd^4x k(z)\hF_{z\mu}^2=\\
  =& -\frac{51N_c}{8960\lambda}\int d^4 x \left[\epsilon^{\mu z \nu_1 \nu_2 \nu_3}\tr\left(R_{\nu_1}R_{\nu_2}R_{\nu_3}\right)\right]^2 \ .
  \end{split}
\end{equation}
The full effective sextic term is then given by $S_6=S^{CS}_{6}+S^{YM}_{6}$:
\begin{equation}\label{fullsextic}
  S_6= \frac{51N_c}{8960\lambda} \int d^4 x \left[\epsilon^{\mu z \nu_1 \nu_2 \nu_3}\tr\left(R_{\nu_1}R_{\nu_2}R_{\nu_3}\right)\right]^2 \ .
\end{equation}

We now want to investigate which are the vector mesons we are integrating out of our action to obtain the sextic term.
In the Sakai-Sugimoto model, the whole tower of vector and axial-vector mesons is included in the field $\mathcal{A}$ as noted in the field expansion (\ref{modesakai}): the functions $\psi_n (z)$ correspond for each value of $n$ to a certain meson.
The defining equation for the functions $\psi_n$ is given by
\begin{equation}\label{mesondef}
  -{h(z)}^{-1}\partial_z\left(k(z)\partial_z\psi_n\right) = \lambda_n \psi_n
\end{equation}
where the eigenvalues $\lambda_n$ determine the mass of the mesons. We can assume that the functions $\psi_n$ are even or odd functions of $z$, since (\ref{mesondef}) is invariant under $z\rightarrow-z$.
If we order the values $\lambda_n$ to be increasing with $n$, then the associated eigenfunctions are of alternate parity, starting with an even $\psi_1$.
From (\ref{modesakai}) we can see that modes that are even (odd) in $z$ correspond to vector (axial) mesons: the lightest vector meson is then to be associated to the function $\psi_1$.

The functions are normalized via the ortho-normality condition
\begin{equation}\label{orthonormalpsi}
  \int dz h(z)\psi_n \psi_m = \delta_{nm} \ .
\end{equation}
So we normalize the function $\chi(z)$ in the same way:
\begin{equation}
|\chi|^2 = \int dz h(z) \chi^2(z)\ .
\end{equation}
\begin{equation}
 \chi^{({\rm norm})}(z)\equiv \frac{\chi(z)}{|\chi|} \ .
\end{equation}
In order to extract the meson content of our solution, we then project the function $\chi (z)$ on the  set of eigenfunctions  via the product
\begin{equation}\label{projectionchi}
 a_n \equiv \int dz h(z)\chi^{({\rm norm})} \psi_n \ .
\end{equation}
The functions $\psi_n$ can be numerically obtained by a shooting method and, since the function $\chi(z)$ is even, we can perform just the projections involving $\psi_{2n-1}$, signifying that we are integrating out only the vector meson tower.
Our results for the squared coefficients $a_n^2$ are the following:
\begin{equation}\label{resultcomponents}
  a_1^2 = 0.988 \ , \quad  \quad a_3^2 = 0.0115  \ ,\quad  \quad a_5^2 = 0.00029 \ .
\end{equation}
We thus see that most of the contribution comes from the $\omega$ meson which is the one associated to the function $\psi_1$.

In the SS model a mass term for the quarks can be introduced via the Aharony-Kutasov action \cite{Aharony:2008an}:
\begin{equation}\label{Sak}
  S_{AK} = c \int d^4 x \tr \,\mathcal{P} \left[M\left(e^{-i\int dz \mathcal{A}^{(\rm bg)}_z}-\mathds{1}\right)+ {\rm c.c.} \right] \ .
\end{equation}
This term breaks the gauge invariance: if we fix a gauge, then we must account for the gauge variation of this term.
The field $\mA^{(\rm bg)}_z$ in this action should be regarded as having the form it had before moving to the gauge $\mA_z=0$ (bg stands for ``before gauge").
We know that the path-ordered exponential of the $\mA_z$ field is to be identified with the pion matrix $\mU$ as follows
\begin{equation}\label{defU}
  \mathcal{P} e^{-i\int dz \mA_z^{(\rm bg)}} = \mU (x) \ .
\end{equation}
Now we take the quark mass matrix to be diagonal, and the masses of the up and down quark to be degenerate and equal to $m$, so we end up with:
\begin{equation}
  S_{AK}= mc \int d^4 x \tr\left[\left(\mU - \mathds{1}\right)+ {\rm c.c.} \right] \ .
\end{equation}
We can now adopt the usual decomposition of the field $\mU$
\begin{equation}\label{decompU}
  \mU \equiv \sigma + i \vec{\pi}\cdot \vec{\tau} \ ,
\end{equation}
together with the unitary constraint
\begin{equation}\label{constr}
  \sigma^2 + |\vec{\pi}|^2 = 1
\end{equation}
The trace the part valued in $SU(2)$ vanishes, while the complex conjugate accounts for a factor of two,
so we end up with the following potential:
\begin{equation}\label{S0}
  S_{0} = 4mc \int d^4 x \left(\sigma - 1\right) \ .
\end{equation}

Remembering that the non-Abelian part of the Yang Mills action produces the quadratic and the quartic term of a Skyrme model (after trivial integrations of the holographic coordinate), we can now write down explicitly the full Lagrangian
\begin{equation}\label{Actionfull}
  S= S_{6} + S_{4} + S_{2} + S_{0}
\end{equation}
with
\bea\label{actiondecomp}
       S_6 &=  & \frac{51N_c}{8960\lambda} \int d^4 x \left[\epsilon^{\mu z \nu_1 \nu_2 \nu_3}\tr\left(R_{\nu_1}R_{\nu_2}R_{\nu_3}\right)\right]^2 \nn \\
     S_4 &=  & a\lambda N_c \int d^4 x \tr\left(\left[R_{\mu},R_{\nu}\right]^2\right)\nn \\
     S_2 &=  &  \frac{\lambda N_c}{216 \pi^4}  \int d^4 x \tr \left(R_{\mu}R^{\mu}\right)\nn \\
     S_0 &=  &  4mc \int d^4 x \left(\sigma - 1\right) \ ,
  \eea
where $a\equiv 1,17 \times 10^{-5}$ and $mc$ fixed by the Gell-Mann-Oakes-Renner relation $4mc = f_{\pi}^2 m_{\pi}^2$.
We are thus left with a generalized Skyrme model provided with the usual pion mass potential $S_0$.

Setting to zero the $SU(2)$ valued vector mesons as we have done in the ansatz (\ref{decomposition})  does not give the full effective Lagrangian: other terms are generated if we integrate out all the vector meson.
Let us comment on this choice by turning on the $\rho$ meson (the same considerations hold for all the vectors).
In this case the new field content reads \begin{equation}
  \mA_{\mu} =
\left\{
  \begin{aligned}
 \hA_{\mu} & = B_{\mu}(x) \chi (z)\\
 A_{\mu} & =R_{\mu}\psi_+ (z) + B^{(1)}_{\mu}(x)\psi_1(z) \ . \\
 \end{aligned}
\right.
\end{equation}
Now we now look at the equation for the field $A_{\mu}^a$ \begin{equation}
  -\kappa \left[h(z)\left(D_{\nu}F^{\mu\nu}\right)^a+\partial_z\left(k(z)F^{\mu z}\right)^a\right] + \frac{N_c}{64\pi^2} \epsilon^{\mu \alpha_1 \cdots \alpha_4} F^a_{\alpha_1 \alpha_2}\hF_{\alpha_3 \alpha_4}=0
\end{equation}
We can again think that in an effective field theory approach, every field will be of some order $L^{-k}$, with the pion field being the leading term (with the lowest value of $k$): if the new field $B^{(1)}_{\mu}$ is of the same order of the abelian vector meson $B_{\mu}$, then it is possible for it to generate quartic and sextic order effective potentials for the pion field. However, if this is the case, all the leading order terms in the equation of motion would come from the Yang-Mills action, resulting in a $\rho$ meson field of order $B^{(1)}\psi_1 \sim \mathcal{O}(\lambda^0)$, while we have $B_{\mu}\chi \sim \mathcal{O}(\lambda^{-1})$. We expect the factorization proposed to be valid in the small $\lambda$ regime, so that the $\rho$ meson contribution to the sextic potential will be suppressed as $\lambda^2$ and not included in our analysis, while we will argue that the quartic potential becomes negligible anyway in both the massive and massless models as $\lambda$ becomes small.

\section{Baryons as Skyrmions }
\label{tre}

\begin{figure}[h]
  \centering
  \includegraphics[width=12cm]{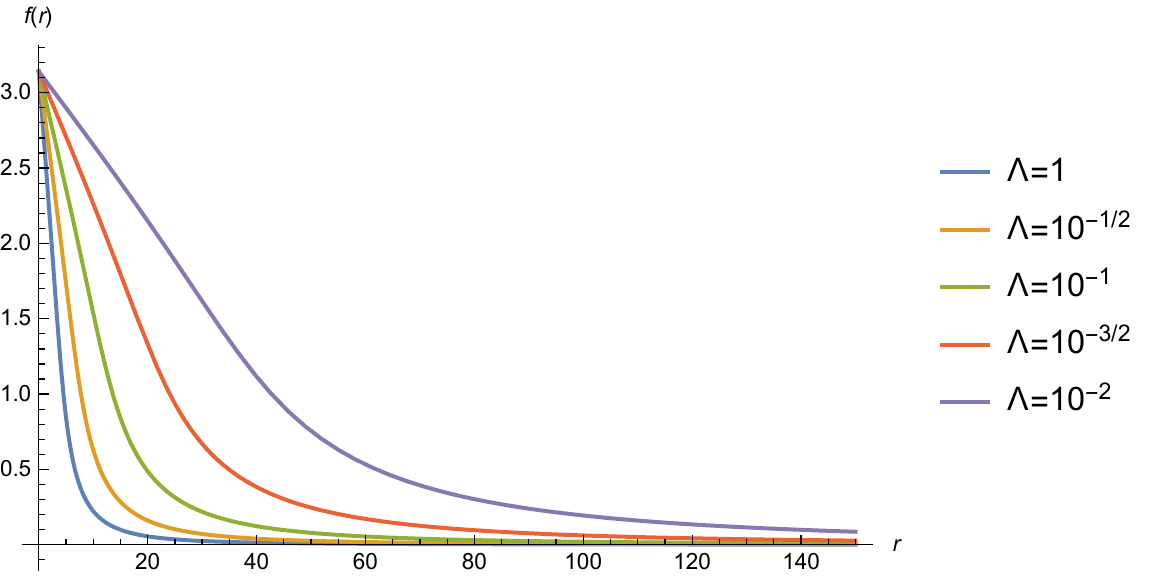}
  \caption{\small Skyrmion profile for decreasing values of $\Lambda$. As can be seen, the size of the soliton solution increases as $\Lambda$ becomes small.}\label{masslesssol}
\end{figure}

We start evaluation the static Action where  we set every time derivative to zero:
\bea\label{staticaction}
  S_{\rm static}  &=&  \frac{\lambda N_c}{216\pi^4}\int d^4 x \tr\left(R_i R_i\right) + a\lambda N_c \int d^4 x \tr\left(\left[R_i , R_j\right]^2\right)  \nn \\
 && \ - \frac{51N_c}{8960\lambda} \int d^4 x\left[\epsilon^{ijk}\tr\left(R_i R_j R_k\right)\right]^2 \ .
\eea
Note that the minus sign in the sextic term comes from $\eta_{00}=-1$ used to contract $\epsilon^{\mu_1\mu_2\mu_3\mu_4}$.  \\
We employ the usual Skyrme model $B=1$ hedgehog ansatz:
\begin{equation}\label{hedgehog}
  \mU(x) = e^{i f(r) \hat{x}\cdot \vec{\tau}} \ .
\end{equation}
The specific form of $f(r)$ will be the one which minimizes the static energy.
Remembering that $S= -\int dt E$, we find:
\bea\label{staticE}
  E  &=&  \frac{\lambda N_c}{27\pi^3}\int dr\left(r^2{f'}^2+2\sin^2f\right) + 64\pi a\lambda N_c \int dr \left(\frac{\sin^4f}{r^2}+2\sin^2f {f'}^2\right) \nn \\ && \ + \frac{459 \pi N_c}{140\lambda}\int dr \frac{{f'}^2}{r^2}\sin^4 f \ .
\eea
From this expression, we can derive the Euler-Lagrange equation for the function $f(r)$:
\begin{equation}\label{ELfunction}
\begin{split}
  \sin 2f&\left\{f'^2\left(b+\frac{\alpha}{\Lambda^2}\frac{\sin^2f}{r^2}\right)-b\frac{\sin^2f}{r}-1\right\}\\
  &+2f'\left\{r-\frac{\alpha}{\Lambda^2}\frac{\sin^4f}{r^3}\right\}+f''\left\{r^2 +2b\sin^2f+\frac{\alpha}{\Lambda^2}\frac{\sin^4f}{r^2}\right\}=0
  \end{split}
\end{equation}
where we have introduced the following parameters:
\begin{equation}\label{newparams}
  \Lambda \equiv \frac{8\lambda}{27\pi} \ , \quad \quad b= 1728\pi^4a\simeq 1.97 \ , \quad  \quad \alpha\simeq76.701 \ .
\end{equation}
The equation has to be solved with the usual boundary values $f(0)=\pi$ and $f(\infty)=0$: this can be achieved numerically by a shooting method, with results showed in Figure \ref{masslesssol} for various values of the $\Lambda$ parameter.

Following Derrick's theorem \cite{Derrick:1964ww}, we expect the size of the soliton to scale as $\Lambda^{1/2}$ in the small $\Lambda$ limit: in fact, rescaling the coordinates as $x^i \rightarrow R x^i$ we can see from (\ref{staticaction}) that the various contributions to the static energy scale like
\begin{equation}\label{derrick246}
  E(R) = \Lambda R E_2  + \Lambda\frac{E_4}{R} + \frac{E_6}{\Lambda R^3}
\end{equation}
Imposing $\frac{dE(R)}{dR}=0$ and the small $\Lambda$ limit we find that $R\sim \Lambda^{-1/2}$, so that in this regime the energy becomes
\begin{equation}
E(R)|_{R=\Lambda^{-1/2}} = \frac{1}{R}(E_2 + E_6) = \Lambda^{1/2}(E_2 + E_6)
\end{equation}
and thus the model should approach a Skyrme model involving only the kinetic term and the sextic term ($\mathcal{L}_{26}$ model).

\begin{figure}[th]
  \centering
  \includegraphics[width=12cm]{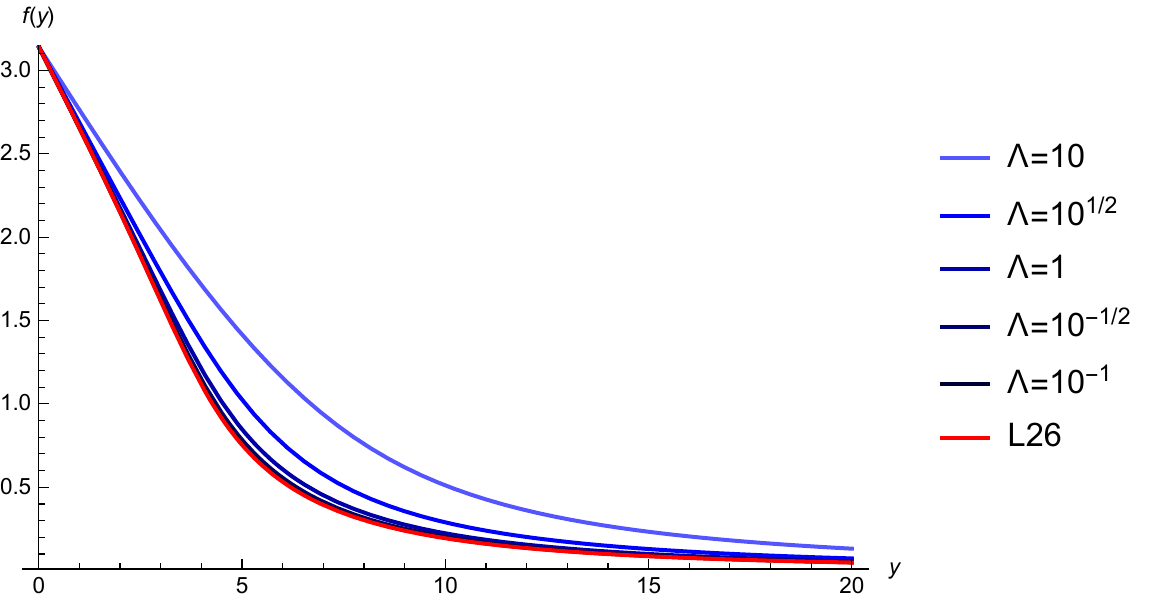}
  \caption{\small Skyrmion's profile functions rescaled to the same size. In red we have plotted the rescaled solution of (\ref{eqn26}), in shades of blue the ones of the full massless model}\label{scale26}
\end{figure}

To explicitly check that this is indeed the case, we rescale the solutions of the massless model adopting a new coordinate $y\equiv r\Lambda^{1/2}$ for various values of $\Lambda$, and we plot the profile functions $f(y)$ together with the one of the $\mathcal{L}_{26}$ model, which can be obtained by solving
\begin{equation}\label{eqn26}
  \sin 2f\left\{f'^2\frac{\alpha}{\Lambda^2}\frac{\sin^2f}{r^2}-1\right\}
  +2f'\left\{r-\frac{\alpha}{\Lambda^2}\frac{\sin^4f}{r^3}\right\}+f''\left\{r^2 +\frac{\alpha}{\Lambda^2}\frac{\sin^4f}{r^2}\right\}=0 \ .
\end{equation}
The results of the numerical solution are plotted in Figure \ref{scale26}: it is manifest that as $\Lambda$ decreases, the quartic term rapidly becomes negligible, and the solution of the $\mathcal{L}_{246}$ model approaches that of the $\mathcal{L}_{26}$ one.
In the large $\Lambda$ limit otherwise we should expect the soliton size to become independent of $\Lambda$ as can be seen again from (\ref{derrick246}): choosing $R \sim 1$ we can again minimize the energy, except that this time the analysis is valid in the large $\Lambda$ region.

\begin{figure}[h]
  \centering
  \includegraphics[width=10cm]{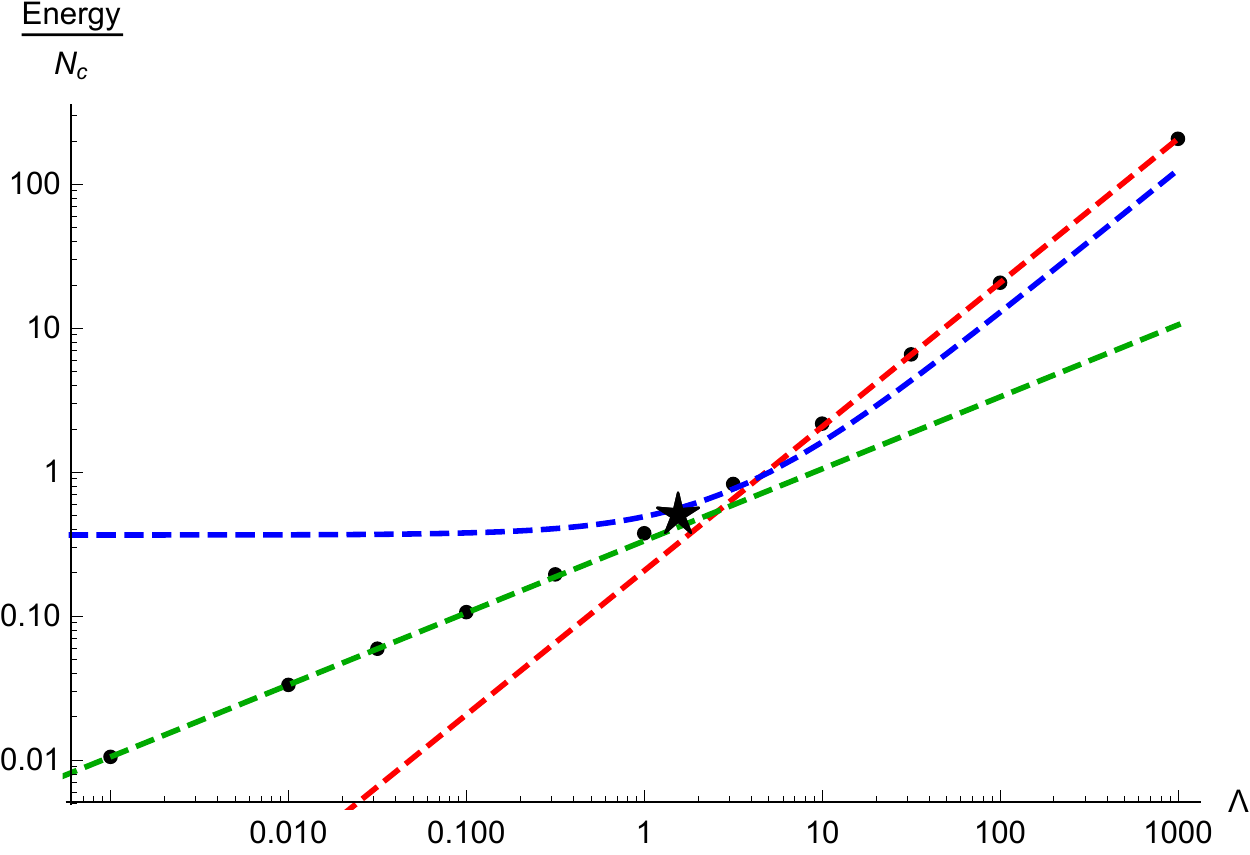}
  \caption{\small Energy for a range of values of $\Lambda$ in the massless model. Black dots are the energy obtained with the model developed. The dashed red and green lines correspond respectively to fitting linear and square root relations, as are expected to develop in the large and small $\Lambda$ regions. The dashed blue line represents the Energy of the BPST instanton which becomes the correct description of the baryon at large $\Lambda$ \cite{Hata:2007mb}. The black star corresponds to the energy computed at the phenomenological value of $\Lambda=1.568$: as can be seen, this value lies just in between the two regimes of large and small 't Hooft coupling.} \label{enmassless}
\end{figure}

Finally, we check these considerations by explicitly computing the total energy for a wide range of values of $\Lambda$, with results shown in the plot in Figure \ref{enmassless}. As can be seen, at small $\Lambda$ the Energy dependence on this parameter correctly approaches the square root one, as predicted by the Derrick's theorem.


It is now interesting to check what happens to the Skyrmion when we consider the full massive model, that is the generalized Skyrme model with pion mass potential, in both the generic and small $\Lambda$ regime.


The new potential $S_0$ modifies the equation of motion for the profile function $f(r)$, which now reads:
\begin{equation}\label{ELmass}
\begin{split}
  \sin 2f&\left\{f'^2\left(b+\frac{\alpha}{\Lambda^2}\frac{\sin^2f}{r^2}\right)-b\frac{\sin^2f}{r}-1\right\}+\\
  &+2f'\left\{r-\frac{\alpha}{\Lambda^2}\frac{\sin^4f}{r^3}\right\}+f''\left\{r^2 +2b\sin^2f+\frac{\alpha}{\Lambda^2}\frac{\sin^4f}{r^2}\right\} - m_{\pi}^2 r^2 \sin f=0 \ .
  \end{split}
\end{equation}
where we have used the explicit form of $f_{\pi}$ given by the holographic model, which can be read from the coefficient of $S_2$ in (\ref{actiondecomp}), and once written in terms of the new parameter (\ref{newparams}) it amounts to:
\begin{equation}\label{fpidef}
  f_{\pi}^2 = \frac{\Lambda N_c}{16\pi^3} \ .
\end{equation}

\begin{figure}
  \centering
  \includegraphics[width=13cm]{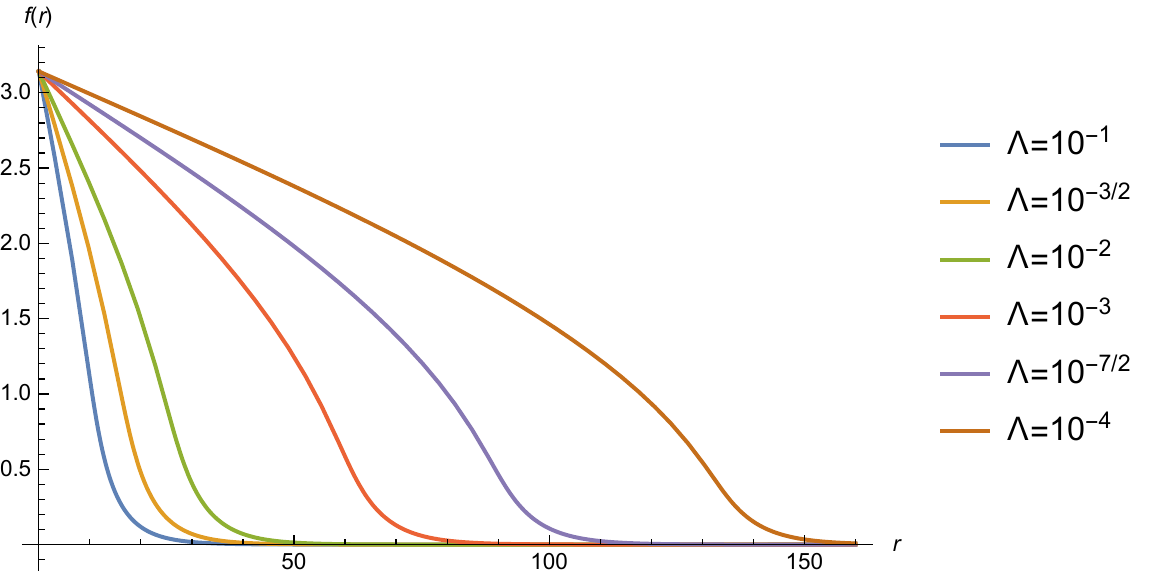}
  \caption{\small Skyrmion's profile functions in the generalized model with pion mass potential.}\label{solmass}
\end{figure}
As before, we can solve (\ref{ELmass}) via a shooting method with the same boundary condition ($f(0)=\pi$, $f(\infty)=0$) to have the $B=1$ soliton. As regards the mass of the pion, we use the phenomenological value of $m_{\pi^0}$, measured in units of $M_{KK}$:
\begin{equation}\label{mpionmkk}
  m_{\pi} \equiv \frac{m_{\pi^0}^{\rm ph}}{M_{KK}} \simeq \frac{135 MeV}{949 MeV} \simeq 0.142
\end{equation}
The plot in Figure \ref{solmass} shows the different shape that the function assumes in the small $\Lambda$ regime, which can be traced back to the model effectively becoming a BPS Skyrme model with only the mass and sextic potentials ($\mathcal{L}_{06}$).

\begin{figure}
  \centering
  \includegraphics[width=10cm]{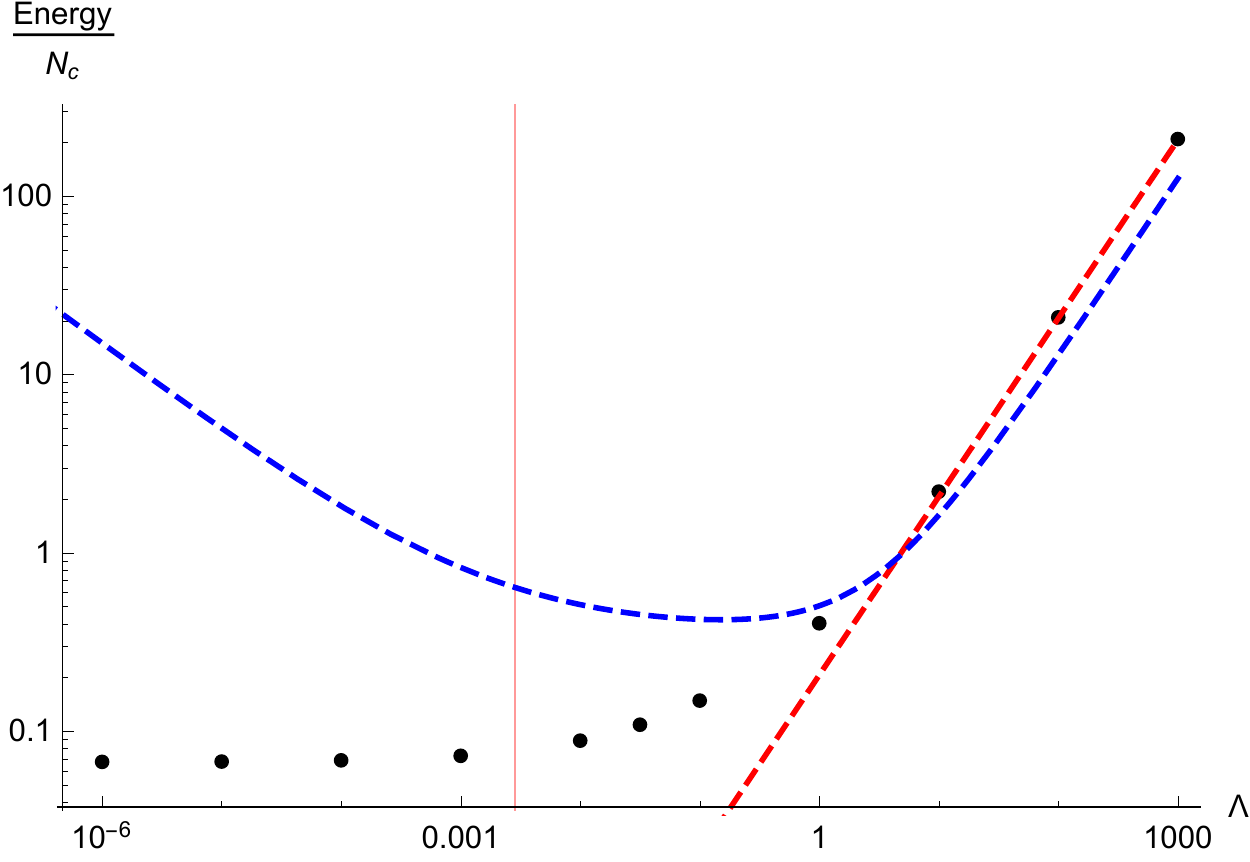}
  \caption{\small Energy for a range of values of $\Lambda$. The dashed red line corresponds to a fitting linear relation, as it is expected to develop in the large $\Lambda$ region. The vertical red line corresponds to the value satisfying $\Lambda = R^{-3} = m_{\pi}^3$, which is expected to be the scale at which the change in behaviour happens. The blue dashed line again represents the Energy of the BPST configuration which describes the baryon at large $\Lambda$, with the addition of the shift in mass computed in \cite{Hashimoto:2009hj} due to the quark mass term.}\label{enmass}
\end{figure}

We can understand this behaviour in the same way as before, observing that this time the soliton size scales as $\Lambda^{-1/3}$ in the small $\Lambda$ regime: after the rescaling the energy contributions scale like
\begin{equation}\label{derrick2460}
  E(R) = \Lambda R E_2  + \Lambda\frac{E_4}{R} + \frac{E_6}{\Lambda R^3} + \Lambda R^3 E_0 \ .
\end{equation}
Minimizing this energy leads again to two possibilities for the size $R$ of the soliton: $R\sim \Lambda^{-1/3}$ or $R \sim 1$.
The latter is again the correct scaling in the large $\Lambda$ regime, and yields to the usual linear dependence of the total Energy on $\Lambda$, while the former is appropriate in the small $\Lambda$ limit and, as can be easily checked from (\ref{derrick2460}), makes the Energy independent of the $\Lambda$ parameter in this limit:
\bea
E(R)|_{R=\Lambda^{-1/3}} &\xrightarrow{\Lambda \rightarrow 0} &\frac{E_6}{\Lambda R^3} + \Lambda R^3 E_0 = E_6 + E_0 \\
E(R)|_{R=1}  &\xrightarrow{\Lambda \rightarrow \infty}& \Lambda (E_2 + E_4 + E_0)
\eea
Both the behaviours are explicitly shown in Figure \ref{enmass}.

The behaviours at large $\Lambda$ are extrapolations reported to compare with the correct description of the baryons in that regime, that is a BPS configuration localized deep in the holographic direction. As can be seen, in both the massive and the massless models, the values of the static energy obtained by this extrapolation show the same power-law dependence on $\Lambda$ as the ones obtained from the actual holographic description, but differ from these by an excess in their normalization.

The reduction to the BPS $\mathcal{L}_{06}$ model is even more manifest if we again plot the rescaled profile functions against the analytical compacton solution of such model: following \cite{Adam:2011zza}, if the Lagrangian reads
\begin{equation}\label{L06}
  \mathcal{L} = \frac{\gamma^2}{24^2} \left[\epsilon^{\mu \nu_1 \nu_2 \nu_3}\tr\left(R_{\nu_1}R_{\nu_2}R_{\nu_3}\right) \right]^2  - \mu^2 \left(1-\sigma\right)
\end{equation}
adopting the hedgehog ansatz (\ref{hedgehog}), the compacton solution is given by
\begin{equation}\label{solutionbps}
  f(r)=\left\{\begin{split}
                 2\arccos\left(Ar\right) & \quad \text{for} \quad r\in \left[0 , A^{-1}\right]\\
                  0 \quad \quad& \quad \text{for}\quad r\geq A^{-1}
              \end{split}
              \right.
\end{equation}
with the inverse length scale being $A= \sqrt[\leftroot{-1}\uproot{2}\scriptstyle 3]{\frac{3\sqrt{2}\mu}{4\gamma}}$.

We can map the parameters $\gamma$ and $\mu$ to the ones of our model just by looking at the action coefficients of $S_0$ and $S_6$, resulting in the identification:
\begin{equation}\label{mapparam}
  \gamma^2 = \frac{\alpha N_c}{8 \Lambda \pi^3} \ , \qquad   \qquad \mu^2 = \frac{\Lambda N_c m_{\pi}^2}{16\pi^3}
\end{equation}
so that the size $A^{-1}$ of the compacton becomes $A^{-1} =\sqrt[\leftroot{-1}\uproot{2}\scriptstyle 3]{\frac{4\sqrt{\alpha}}{\Lambda m_{\pi}}} $. Note that, as expected, the size of the compacton scales as $\Lambda^{-1/3}$.
\begin{figure}
  \centering
  \includegraphics[width=13cm]{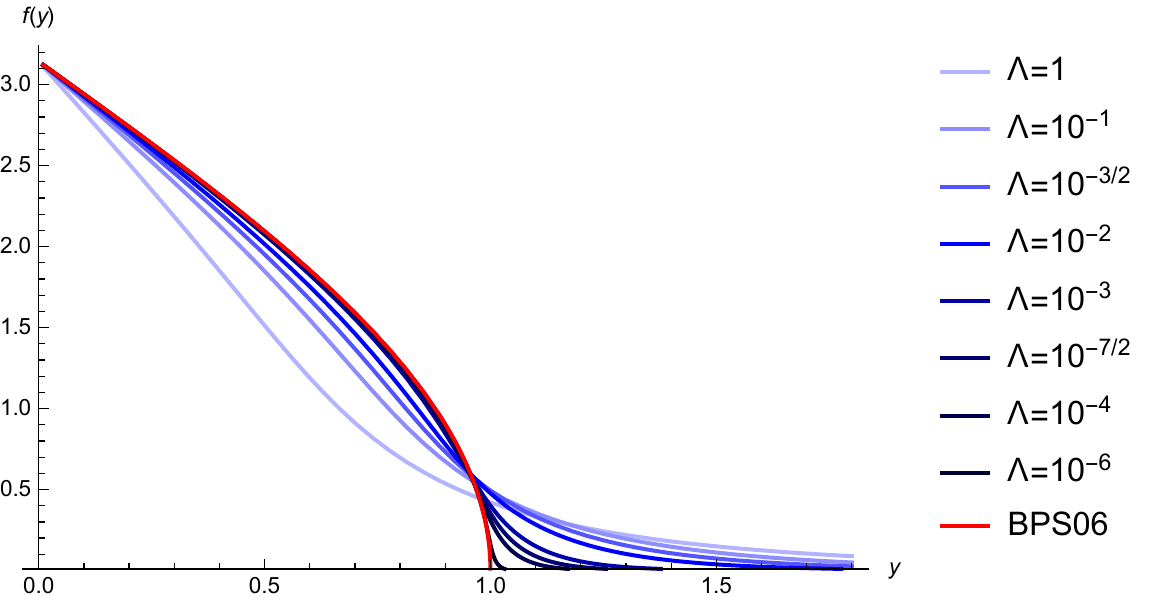}
  \caption{\small Skyrmion's profile functions rescaled to the same size. In red we have plotted the rescaled analytical solution of the BPS model $\mathcal{L}_{06}$, in shades of blue the ones of the full massive equation (\ref{ELmass}).}\label{solscaledmass}
\end{figure}
In Figure \ref{solscaledmass}, we have plotted the solutions of the full equation (\ref{ELmass}), rescaled to the same unitary size via the redefinition $y= Ar$, for a wide range of values of $\Lambda$. As can be seen, the profile functions $f(r)$ approach the analytical one for decreasing $\Lambda$, but not as fast as in the massless case examined before: to reach a good match, we have to push $\Lambda$ way down to values of at least $10^{-6}$, while the massless solution was already in great agreement with the one of the $\mathcal{L}_{26}$ model for $\Lambda=10^{-1}$.

\section{Conclusion}
\label{quattro}

The Sakai-Sugimoto model at low energies may be written as a derivative expansion in the pion fields.
In particular we computed the sextic term which is generated by an integration over the $\omega$ meson and the whole mesonic tower on top of it.
This term becomes particularly important in the small 't Hooft coupling limit.
We also conjectured that the Instanton-baryon, as the 't Hooft coupling becomes small, is well approximated by the Skyrmion computed from the low energy model, which is a generalized version of the Skyrme model. When the pions are strictly massless the dominant terms are $ {\cal L}_2 +  {\cal L}_6$. If pions are massive we recover  the BPS Skyrme model $ {\cal L}_0 +    {\cal L}_6$.

Phenomenological calibration of the SS model requires a choice of the 't Hooft coupling that in general is neither too big or too small. Up to now only the very large 't Hooft coupling region has been analytically solved by the fact that instanton become almost self-dual. We showed here that there is another region in the parameter space where great simplification occurs for the baryons, that of small $\lambda$.  This may help in understanding better the solution in the intermediate regime which, so far, can be accessed only with numerical methods, for the moment \cite{Bolognesi:2013nja}.

Skyrme models in general have always been plagued by the problem of predicting too large classical binding energies.
Quantum effects can help reducing the binding energies, but in general is always better to start from something already close to the real value at the classical level.
In this respect it is quite interesting to have a model that is never too far from a BPS sub-structure. The SS model with the massive deformation has exactly this feature, for big 't Hooft coupling $\lambda$ becomes the self-dual Yang-Mills plus small corrections, while for small $\lambda$ becomes the BPS Skyrme model plus small corrections.  It remains to be seen if this feature can help in reducing, and how much,  the classical binding energy also in the intermediate region of the parameter space.

\section*{Acknowledgments}

We thank C.~Adam, I.~Basile, P.~Sutcliffe, A.~Wereszczynski, for useful discussions.  The present  work
is supported by the INFN special research project grant GAST (``Gauge
and String Theories'').

\end{document}